\documentclass[12pt]{article}
\usepackage[left=2.5cm,right=2cm,top=2cm,bottom=2cm]{geometry}
\usepackage[utf8]{inputenc}
\usepackage[T1]{fontenc}
\usepackage{lmodern}
\usepackage{amsmath,amsfonts,amssymb}
\usepackage{empheq}
\usepackage{indentfirst,tocloft,blindtext,caption}

\usepackage{graphicx}
\graphicspath{{./Figs/}}
\usepackage{wrapfig}
\usepackage{subfig}
\usepackage{fancyhdr}

\usepackage{floatrow}
\usepackage[none]{hyphenat}
\emergencystretch=\maxdimen
\begin{document}

\title{Optical signal-based improvement of individual nanoparticle tracking analysis}
\author{Minh-Chau Nguyen*, Pierre Bon * \\ Université de Limoges, CNRS, XLIM, UMR 7252, F-87000 Limoges, France \\ minh-chau.nguyen@cnrs.fr, pierre.bon@cnrs.fr}
\date{}
\maketitle

\vspace{10pt}

\begin{abstract}
Nanoparticle Tracking Analysis (NTA) provides a simple method to determine individual nanoparticle size. However, because size quantification is based on the slowly converging statistical law of random event, its intrinsic error is large, especially in case of limited event number, \textit{e.g.} for weak scattering nanoparticles. Here, we introduce an NTA improvement by analyzing each individual NP trajectory while taking into account the other trajectories with a weighting coefficient. This weighting coefficient is directly derived from the optical signature of each particle measured by quantitative phase microscopy. The simulations and experimental results demonstrate the improvement of NTA accuracy, not only for mono-disperse but also for poly-disperse particle solutions.
\end{abstract}

%
\vspace{2pc}

%
%
%
\newpage
\section{Introduction}

The Brownian motion of a nanoparticle (NP) suspended in solution obeys a statistical law in which the mean square displacement (MSD) depends on the NP size \cite{Einstein_1956}. Nanoparticle Tracking Analysis (NTA) is therefore a common approach for size quantification, not only for individual inert NPs but also nano-size biological objects \cite{NTA_protein,NTA_EVs,NTA_Virus}.  For a NP diffusing in a solution of dynamic viscosity $\eta$, its diameter $d$ can be derived from its 3D MSD following:
\begin{eqnarray}
\centering
\text{MSD} &= \frac{1}{N-1} \sum_{i=1}^{N-1 } (r_{i+1} - r_{i})^2 = 6\, \frac{k_B \, T}{3 \pi \, \eta \, d} \, \Delta t \nonumber \\ 
\Rightarrow \, d &=  \frac{2\, k_B \, T}{ \pi \, \eta  \cdot \text{MSD}} \, \Delta t
\label{eqn:MSD_Size} 
\end{eqnarray}
, where $N$ number of tracked points, $r_i$ particle position at instant $i$, $k_B$ Boltzmann constant, $T$ absolute temperature, $\Delta t$ time lag (duration between two consecutive points). Theoretically, the MSD is proportional to time lags $\Delta t$. Currently, the most accurate method for size estimation relies on a linear fit of first few points of MSD curve as a function of time lags \cite{QianHong_1991,Michalet_2010}. \\

However, this NTA approach possesses a large intrinsic error margin which depends primarily on the length of tracking trajectory (convergence in $1/\sqrt{N}$) and the localization error \cite{QianHong_1991,Michalet_2010,Saxton_1997}. In the case of small nanoparticles, their weak signals lead to an important size quantification inaccuracy, due to both low localization precision and limitation of tracking length. As an example, the relative size error (considering only the limited track length) is estimated about 70\% if the nanoparticle can be followed in 10 consecutive frames and falls to 20\% for 100 frames \cite{Michalet_2010}. As an attempt to improve NTA accuracy, various methods are proposed, either increasing the tracking length (\textit{e.g.}, using a holographic approach in which a nanoparticle can be tracked although positioned far from the imaging plane \cite{Verpillat_2011,Vitor_JACS_2016,Quidant_2023}), or using the localization error as a weighting factor for size estimation \cite{Michalet_2010}. Besides, if particle population statistics are sufficient, multiple post-processing algorithms have been proposed in order to estimate the mean size of solution from the size distribution by using a covariance-based estimator \cite{Flyvbjerg_2014_Covariance_estimator,Cho_2019_NTA_correction, Cho_2019_NTA_Radius}, or maximum likelihood approach \cite{Saveyn_2010_NTA_Likelihood,Walker_2012_Improved_NTA,Matsuura_2017,Silmore_2019}.\\

In this paper, we introduce our individual NTA improvement approach based on the exploitation of the optical signature of each particle and trajectories in an acquisition. Instead of a simple linear fit of all MSD curves (allowing to derive the average size of all the particles but not individual particle \cite{Michalet_2010}), we suggest using weighted linear fit, in which each single particle's MSD curve will not have the same impact for the final size determination of the considered particle. The weighting coefficient represents the optical similarity between particles. It should be in between 0 (extreme case where two particles are considered as totally different) and 1 (in the case of two identical particles).\\

In order to calculate the similarity between particles, quantitative phase microscopy via wavefront imaging is used to obtain two images (intensity and phase) for each particle. These images characterize respectively the absorption/attenuation and the refractive index of the NP with a signal also linked to the particle volume. It is therefore a good criterion for the similarity between particles because it is able to discriminate particles both by their size and nature. To jointly take into account all the optical signatures into the calculation, these two images, intensity ($I$) and phase ($\varphi$), are merged into a unique observable, called Rytov field : 
\begin{equation}
E_{Ry} =  \, \frac{i \, \lambda \, n_m}{\pi} \left[ \frac{\ln (I)}{2} + i \, \varphi \right] \, ,
\end{equation}
where $\lambda$ illumination wavelength, $n_m$ refractive index of the medium. The Rytov field is a complex quantity: its real part (imaginary part, respectively) represents the refraction (absorption/attenuation, respectively) properties of the particle. Moreover, the Rytov amplitude image, $A_{Ryv} = |E_{Ry}|$, exhibits a signal on a zero background for any type of particle, and looks very similar to a fluorescent NP image. This simplifies the tracking algorithm: an universal tracking algorithm can be applied without \textit{a priori} knowledge of the particle optical response \cite{MCN_2023_PhaseVirus}. In addition, Rytov amplitude indicates the interaction between light and NP and is proportional to the number of detected photons. For our application, Rytov field is therefore much more convenient than classic scalar electromagnetic field.\\

We interpret the similarity between two particles from the difference between their Rytov field images. Because of the imaging and sampling condition, Rytov field image of a particle is considered as a sum of particle's actual signal $E_{Ry}^s$ and a noisy background of variance $\varepsilon$. The signal-to-noise ratio (SNR) is defined as $|E_{Ry}^s|/\sqrt{\varepsilon}$ and is proportional to square root of the number of detected photon. The variance of the difference between the Rytov field of $i^{th}$ particle  ($E_{Ry,i}$) and the Rytov field of $j^{th}$ particle ($E_{Ry,j}$) is the actual signal variance plus twice the acquisition error (supposing the two particle images have the same noise statistics). We define the similarity $C_{i,j}$ by the following formula: 
\begin{equation}
C_{i,j} =  \frac{\,2 \varepsilon}{\sigma^2(E_{Ry,i}-E_{Ry,j})}  = \frac{\, 2\varepsilon}{\sigma^2(\Delta E_{Ry}^s) + \,2 \varepsilon} \, ,
\end{equation}

where $\sigma^2(E_{Ry,i}-E_{Ry,j})$ the variance of the difference between two Rytov field, $\Delta E_{Ry}^s$ the actual signal difference between two Rytov field in absence of noise. In case of identical particles, the true signal difference is 0 leading to a similarity of 1. When two particles are different, the similarity $C_{i,j}$ is less than 1. The weighting coefficient is defined as an exponentiation of the similarity $C_{i,j}$ by a positive exponent $n_w$:

\begin{equation}
W_{i,j} =  C_{i,j}^{n_w} = \left( \frac{2 \, \varepsilon}{\sigma^2(E_{Ry,i}-E_{Ry,j})} \right)^{n_w} .
\label{eq:def_weight_coeff}
\end{equation}

Since the base $C_{i,j}$ is smaller than 1, using an exponentiation by $n_w$ kepdf the final value between 0 and 1 (1 when two particles are identical, and tends to 0 when the Rytov field difference is important). This exponent $n_w$ is introduced to modulate the averaging process: the arithmetic average is achieved at $n_w = 0$ (all the trajectories are analyzed equally in order to retrieve the average size of particles in solution), while $n_w \to \infty$ corresponds to the classic NTA process in which each particle size is calculated only from its trajectory.\\

\begin{figure*}[h!]  
	\centering
	{\includegraphics[width=1\textwidth]{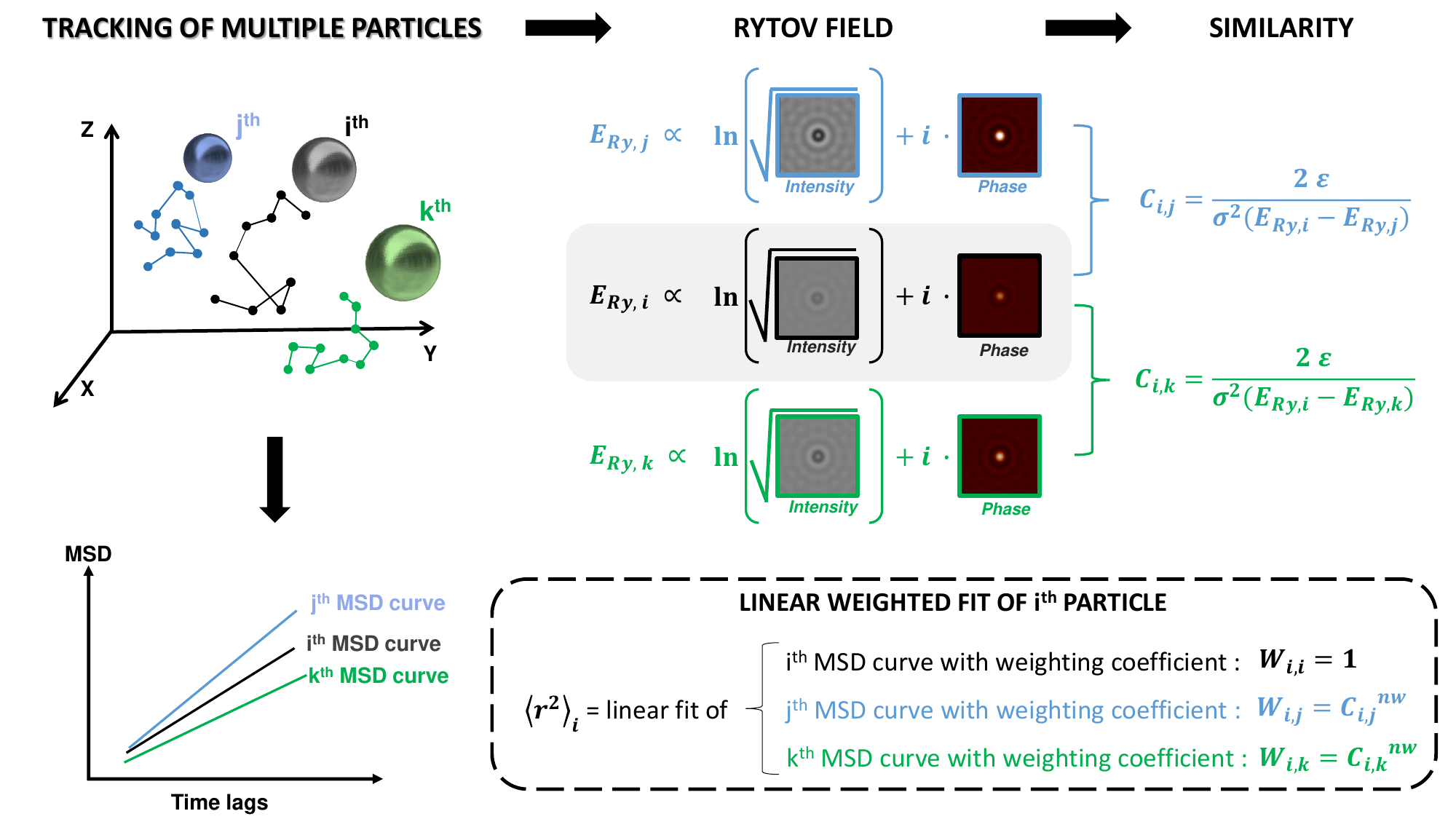}}\hfill
	\caption{\textbf{Principle of method}. Size quantification of $i^{th}$ nanoparticle from all MSD curves.}
	\label{Principle}
\end{figure*}

Figure \ref{Principle} summarizes the principle of our algorithm. 

\section{Results - Discussion}

\subsection{Validation by simulation studies}

In order to study the effect of our weighted MSD fit, we carried out simulation studies (LabView, National Instruments). Particles with sizes satisfying a normal distribution with known mean size and standard deviation (S.D.) are considered and their 200-point Brownian trajectories are generated for each particle. Intensity and phase images of each particle at each position are simulated, using Product-Of-Convolution \cite{Sierra_2009,Bon_2012_modeling}. Camera's shot noise can be added in order to replicate the images in different experimental conditions. \\

From these 200 noise-realistic images, super-localization is applied on the Rytov amplitude to recenter the images although the particle is moving (carried out with realistic simulated localization error of 30-nm along the lateral direction and 100-nm along axial direction). Different averaged sub-images are then produced from the pool of numerically centered images. First, the all 200 sub-images are average to produce an dynamic averaged Rytov field $E_{Ry,i}$. Then, the image dataset is split in $2\times100$ images to produced, via averaging, two images of the same particle $E_{Ry,i,1}$ and $E_{Ry,i,1}$ which differ only by their noise. By computing $\sigma^2(E_{Ry,i,1} - E_{Ry,i,2}) = 2\,\epsilon_i$, one can extract the actual noise amplitude of the particle image $E_{Ry,i}$. Noteworthy, this method can be applied on actual acquisition and not only simulated particles to measure the intrinsic noise level \cite{MCN_2023_PhaseVirus}.\\

\begin{figure*}[h!]  
	\centering
    \subfloat[]
	{\includegraphics[width=0.35\textwidth]{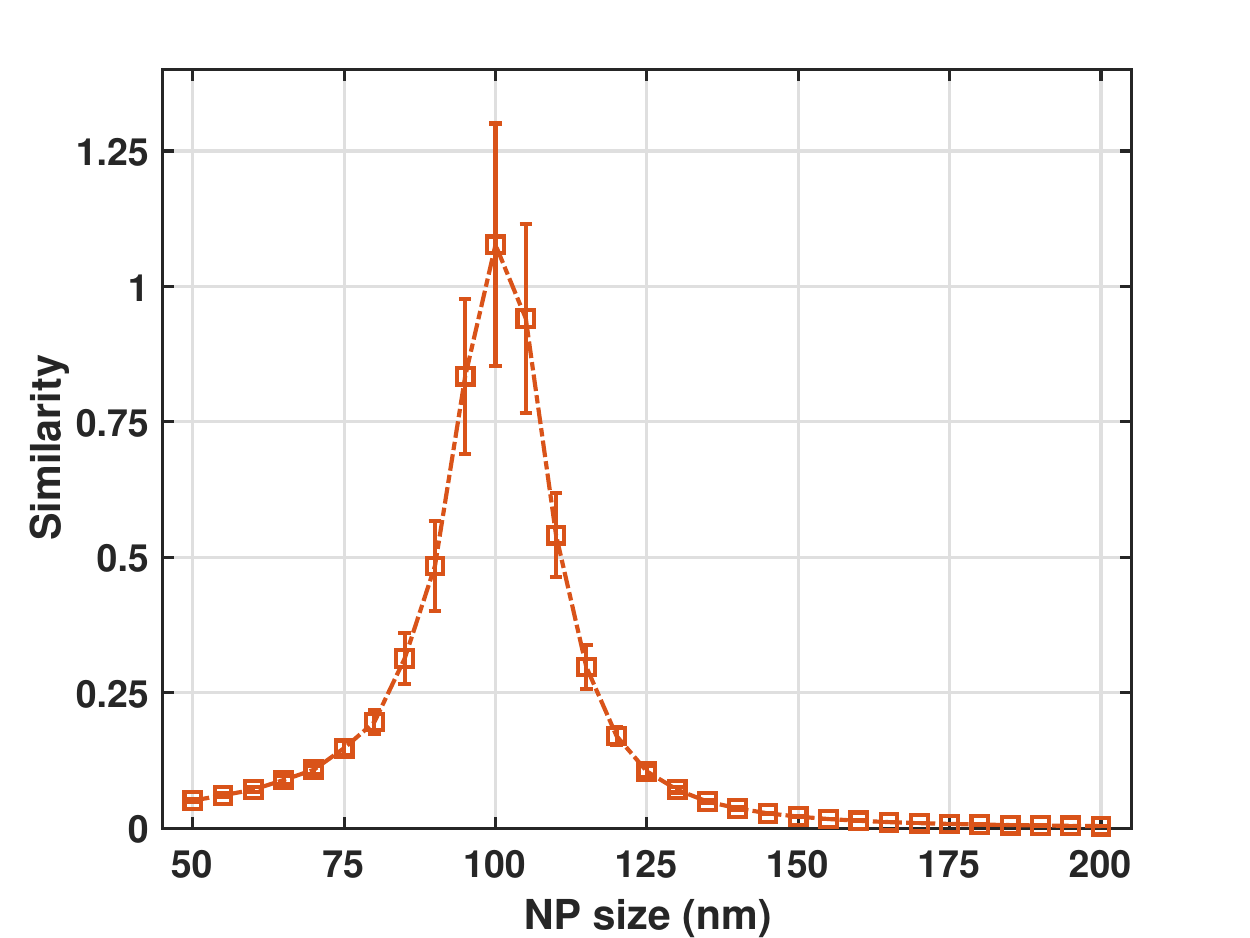}\label{fig:Simul_01_Similarity_vs_Size}}
    \subfloat[]
	{\includegraphics[width=0.35\textwidth]{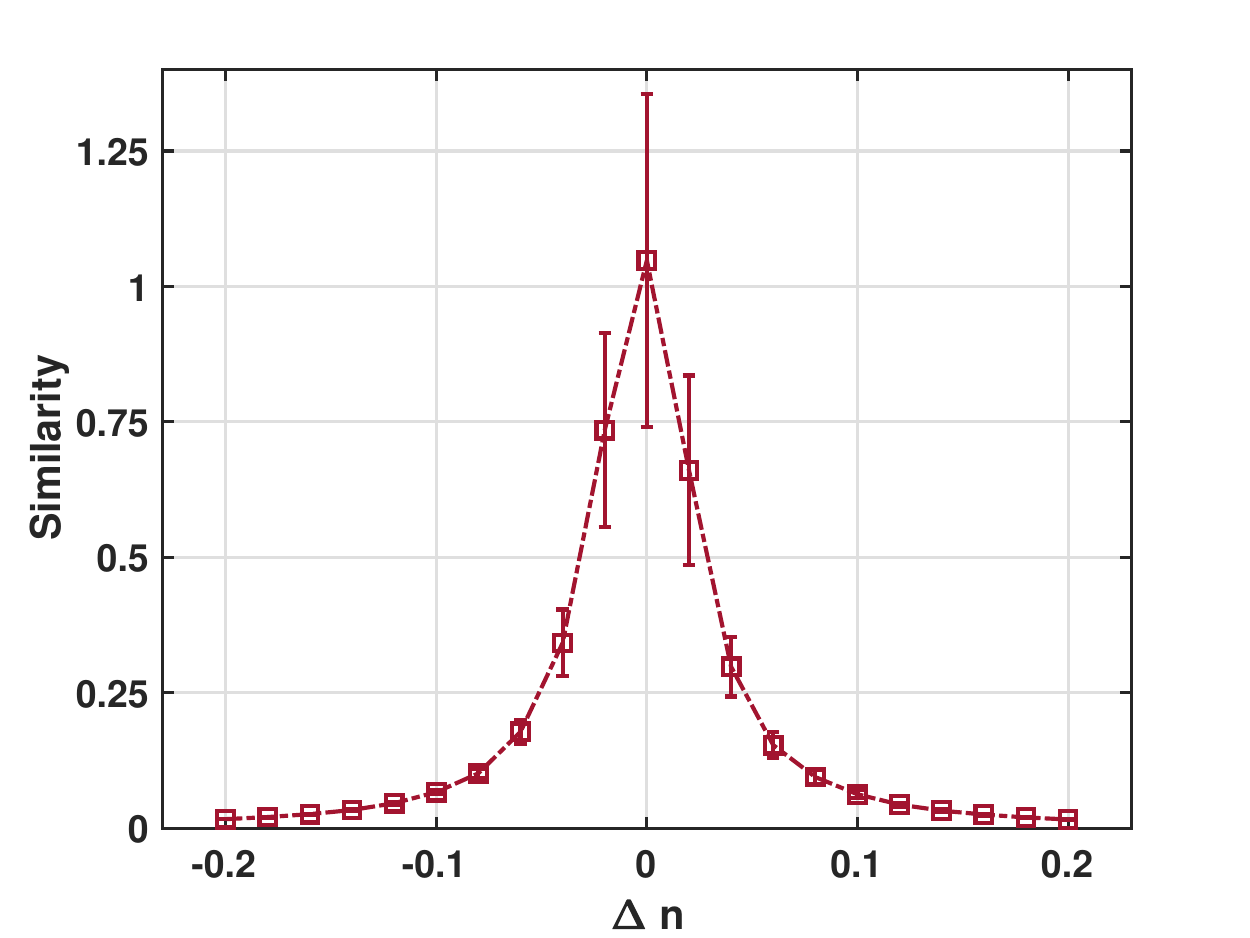}\label{fig:Simul_01_Similarity_vs_Delta_n}}
    \subfloat[]
	{\includegraphics[width=0.35\textwidth]{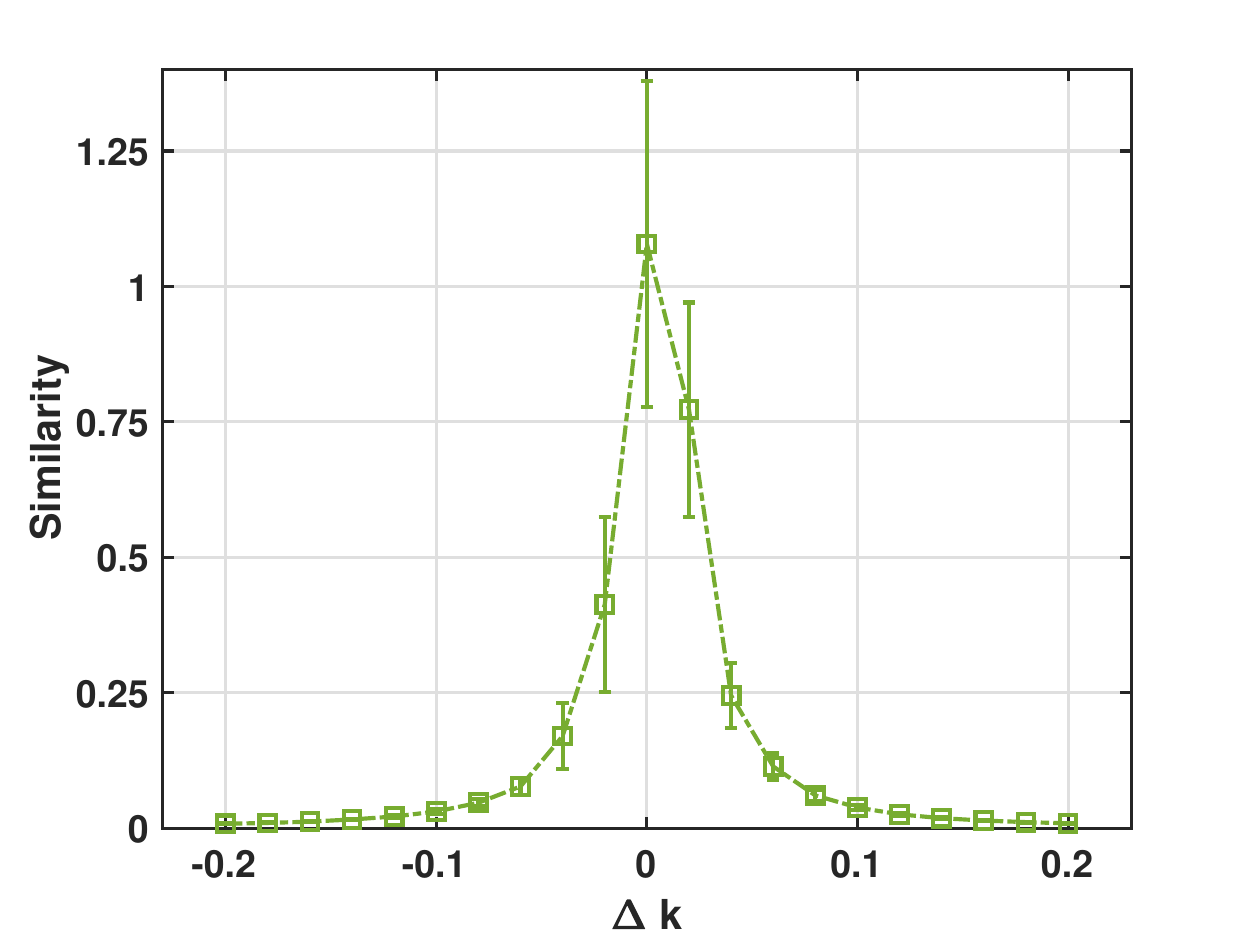}\label{fig:Simul_01_Similarity_vs_Delta_k}}
	\caption{\textbf{Simulation of the similarity between particles.} Similarity between simulated 100-nm PS NP: \textbf{(a)} versus different size PS NP, \textbf{(b, c)} versus 100-nm NP of different refractive index ($n$) and absorption coefficient ($k$). $\Delta n$ and $\Delta k$ compared to the complex refractive index of PS. } 
	\label{Similarity}
\end{figure*}

The study is first carried out for 100-nm polystyrene (PS) NP. Figure \ref{Similarity} depicts the variation of similarity between a simulated 100-nm PS NP and with different particles, while varying NP size, or NP complex refractive index $\tilde{n}= n + i\cdot k$. The similarity reaches 1 when the particle is compared to another particle having same characteristics, and starts vanishing quickly with the difference in size (and/or refractive index, and/or absorption coefficient). The similarity coefficient can surpass 1 in this simulation for 2 exactly (or quasi) similar particles due to the acquisition noise. If the similarity is higher than 1, its value is set at 1.

\begin{figure*}[ht!]  
	\centering
    \sidesubfloat[]
	{\includegraphics[width=0.45\textwidth]{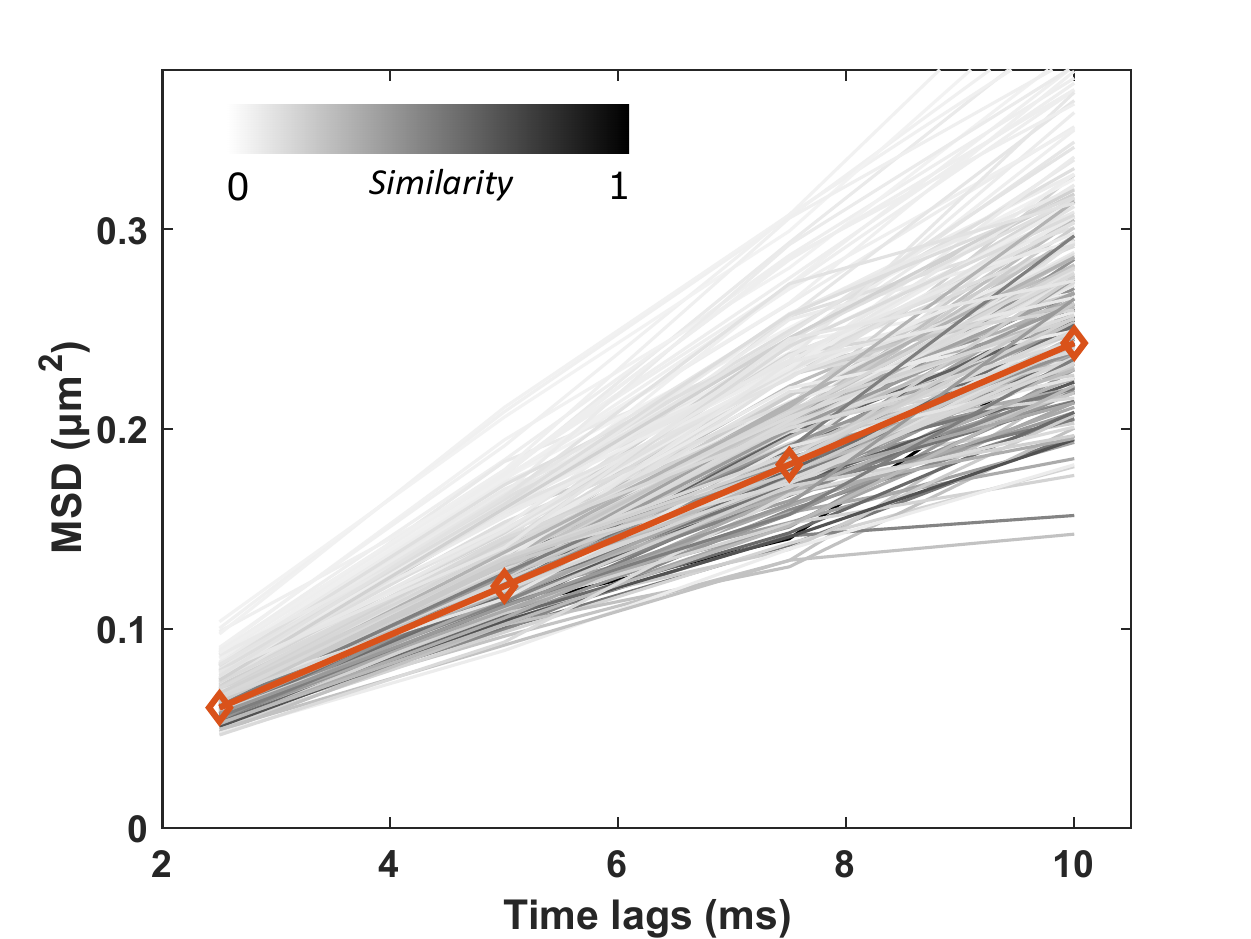}\label{fig:MSD_all}} 
    \sidesubfloat[]
	{\includegraphics[width=0.45\textwidth]{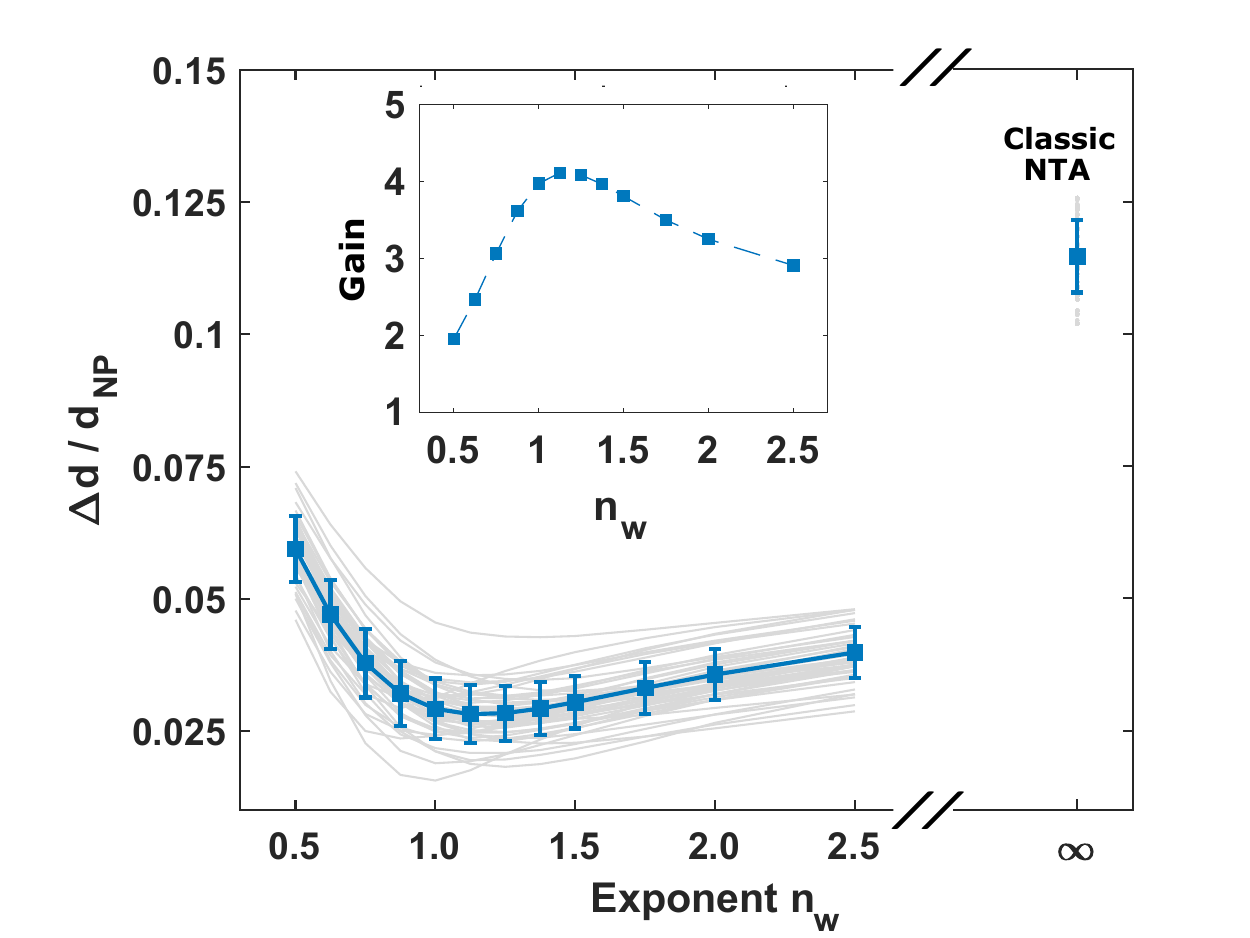}\label{fig:Simul_02_Error_vs_nw}}\linebreak
	\sidesubfloat[]
	{\includegraphics[width=0.95\textwidth]{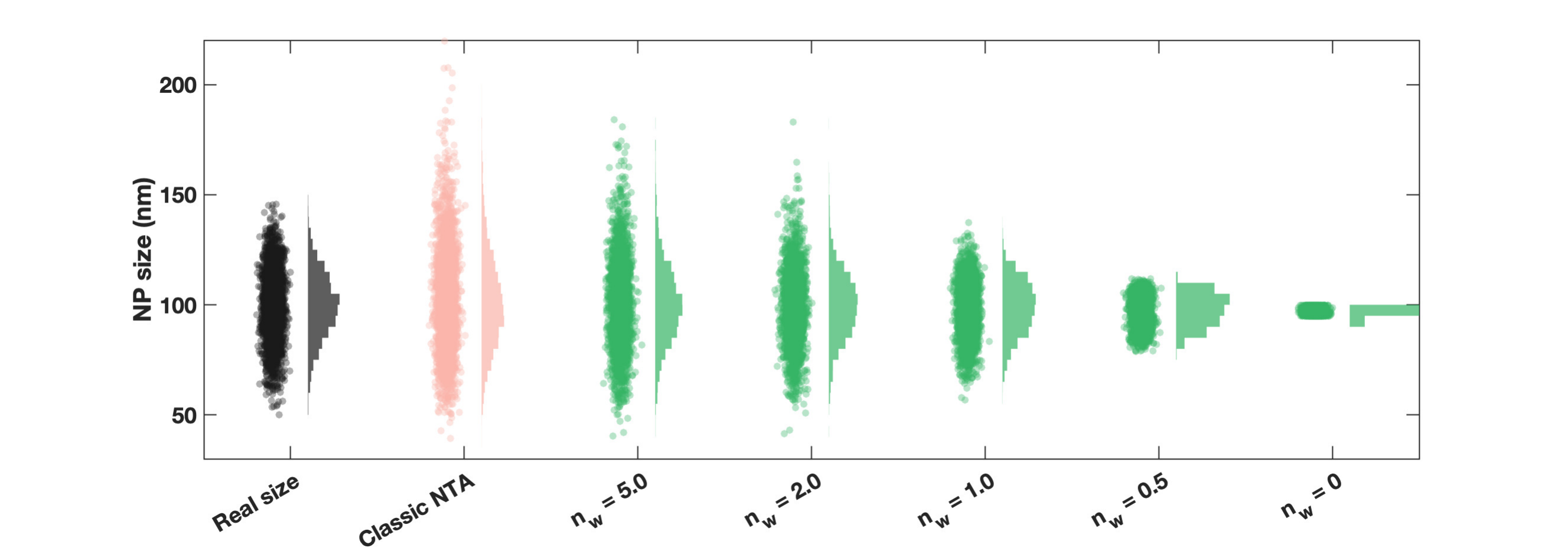}\label{fig:Simul_Fig_2_Histo_nw}} \linebreak
    \subfloat[]
	{\includegraphics[width=0.35\textwidth]{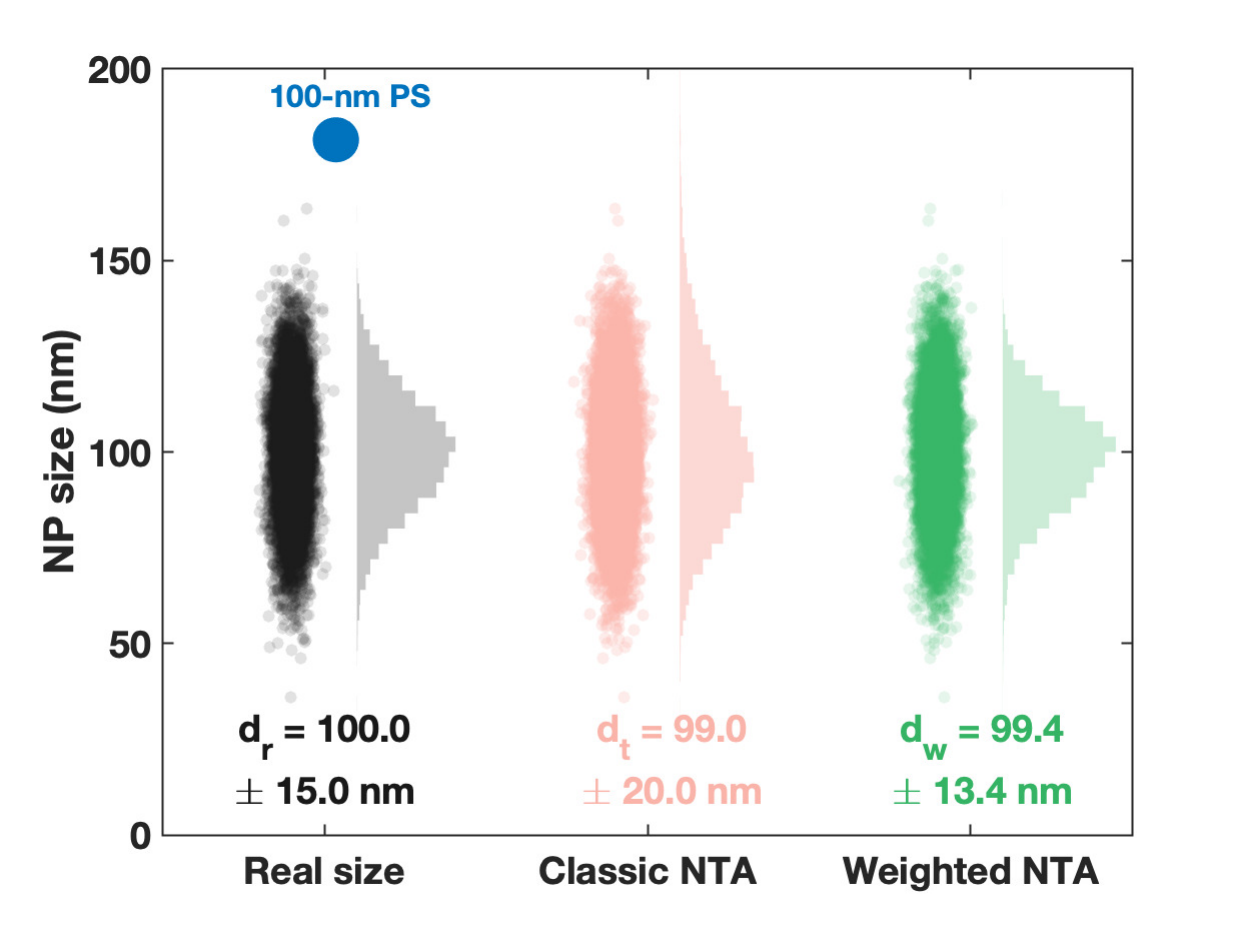}\label{fig:Simul_Fig_2_Histo_PS100}}
     \subfloat[]
	{\includegraphics[width=0.35\textwidth]{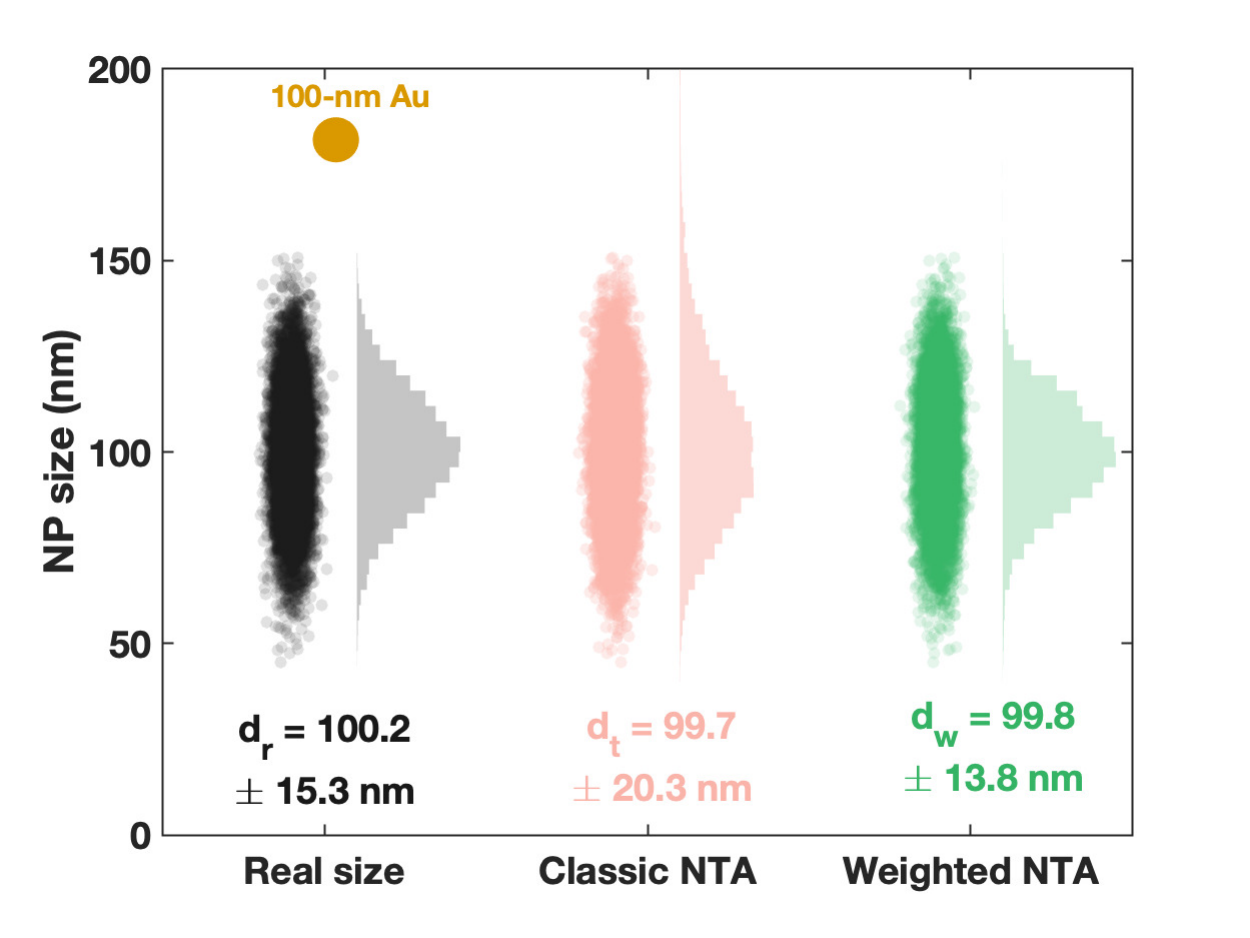}\label{fig:Simul_Fig_2_Histo_Au100}}
    \subfloat[]
	{\includegraphics[width=0.35\textwidth]{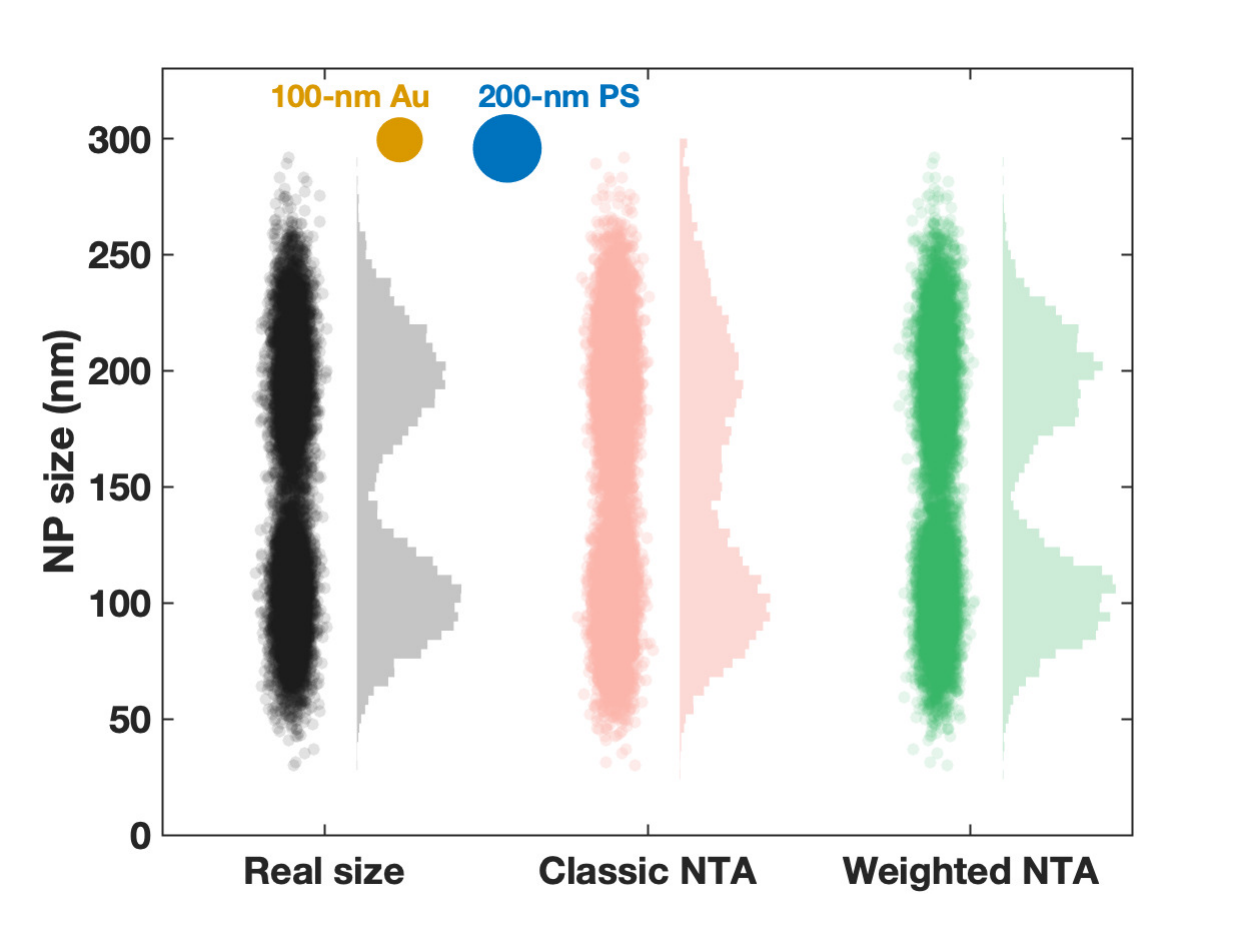}\label{Simul_Fig_2_Histo_Mix}}
	\caption{\textbf{Simulation of MSD curve's weighted fit}. \textbf{(a)} Example of the MSD curve's weighted fit of one particle versus the MSD curves of all the particles. \textbf{(b)} Relative size difference in function of the exponent $n_w$ for 50 repetitions (gray line) and its statistics (mean value and standard deviation, in blue line and points). Exponent $n_w = \infty$ (on the top right) represents the classic NTA. Inset graph illustrates the gain in NP sizing as a function of $n_w$. \textbf{(c)} Distribution of NP sizes of a 100-nm PS solution calculated by weighted fit at different exponents, compared to classic tracking analysis, and its real size. \textbf{(d, e, f)} Comparison of the distribution of the real size, the classic NTA and the weighted NTA for 3 solutions : 100-nm PS NP (d), 100-nm Au NP (e), and a mix of PS and Au NP (f).}
	\label{Simul}
\end{figure*}

A second study has been performed for a virtual PS batch of 200 NPs so that their real size are normal distributed with a mean size of 100-nm and a S.D. of 15-nm. Images of each particle are generated without the shot noise (SNR estimated at 55, due to the interference fringes). Figure \ref{fig:MSD_all} illustrates all the 200 MSD curves (gray line), versus the weighted fit MSD curve of the first particle (orange line). The gray scale of each MSD curve is color-coded from the similarity between each particle and the first particle (fader the line color, lower the similarity).\\

Then, we vary the exponent $n_w$ to discuss the effect of the weighted MSD fit process for size retrieval. Let us consider $d_{NP}$ as the actual particle size, $\Delta d$ the difference between the measured size via weighted linear fit and actual particle size, and $\Delta d_{NTA}$ the difference between the measured size via classic NTA and actual particle size. The optimal weighting coefficient is reached when the relative size difference $\Delta d /d_{NP}$ is minimized which corresponds to a maximization of gain$ = \Delta d_{NTA}/\Delta d$. Results are presented in figure \ref{fig:Simul_02_Error_vs_nw} (50 repetitions were carried out, all the repetitions in gray line, its mean value and S.D. in blue line) and we determined that the best $n_w$ exponent was 1.125. Comparing to the classic NTA (correspond to $n_w = \infty$ in the graph), the relative size difference is reduce from 12\% to 3\%, illustrating a gain in NP sizing of about 4 (see inset graph).\\

We have also studied the distribution of NP size determined by the weighted fit at different exponents, compared to the real size and the classic NTA, as illustrated in Figure \ref{fig:Simul_Fig_2_Histo_nw}. When the averaging process is more important (\textit{i.e.} smaller $n_w$ exponent), the dispersion of NP size is smaller than the real dispersion and can be considered as an artifact. In the extreme case where $n_w = 0$, all the trajectories are taken into account equally to extract the average NP size. In the contrary, when $n_w$ tends to infinite ($n_w \, > 5$), the averaging process is vanished and we obtain the distribution of the classic NTA.\\

The histograms of real size, size measured from classic NTA and with the optimal weighted fit (here, $n_w = 1.125$) of our simulated 100-nm PS solution are illustrated in Fig \ref{fig:Simul_Fig_2_Histo_PS100}. The measurement mean size and dispersion in standard deviation are $99\pm 20.0$-nm in the case of classic NTA and $99.4\pm 13.4$-nm in the case of our weighted fit, closer to the size distribution dispersion of $100\pm15$-nm. The weighted fit clearly improves the NP size determination, as its histogram is almost identical to the real histogram. Similar results are obtained for absorbing particles, here 100-nm Au NP (illustrated in Figure \ref{fig:Simul_Fig_2_Histo_Au100}), confirming the interest of using weighted NTA approach.\\

The method is also efficient for poly-disperse solutions. We have considered a mix of 100 particles of PS (mean size of 200-nm, dispersion 25-nm) and 100 particles of gold (Au) NP (mean size of 100-nm, dispersion 20-nm). These particles has been chosen since they have mostly the same SNR in Rytov amplitude images \cite{MCN_2023_PhaseVirus}. The histograms of its real size, classic and weighted NTA, shown in Figure \ref{Simul_Fig_2_Histo_Mix}, also describe an objective improvement of size quantification. For each population of particles, the size dispersion is clearly reduced, from 39.6 to 24.0-nm for PS NP and from 25.7 to 19.5-nm for Au NP while keeping an accurate mean size value (199.3-nm and 99.7-nm for PS NP and Au NP respectively). \\ 

\begin{figure*}[ht]  
    \centering
    \sidesubfloat[]
	{\includegraphics[width=0.45\textwidth]{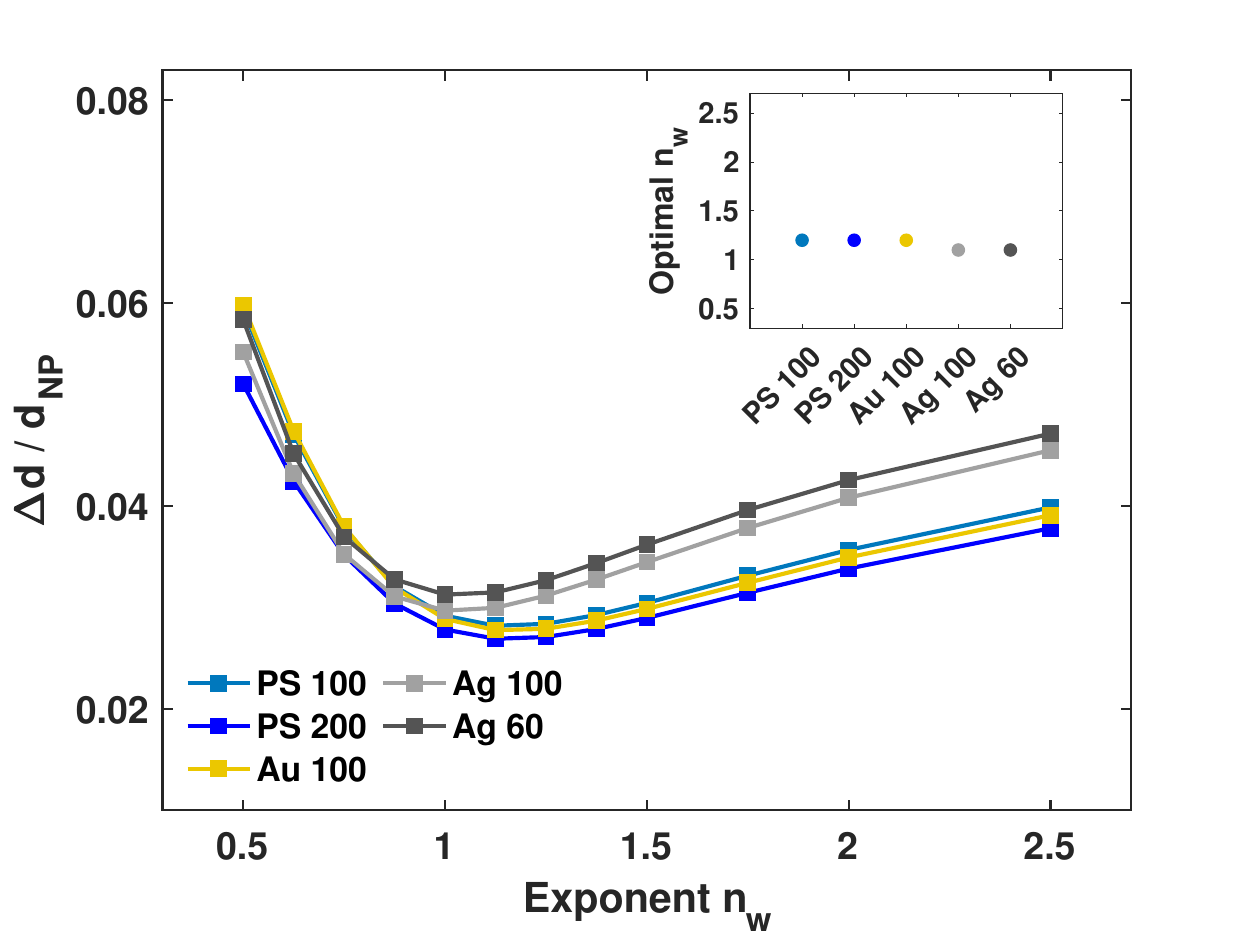}\label{fig:Simul_Fig_2_Diff_Particles}}
    \sidesubfloat[]
	{\includegraphics[width=0.45\textwidth]{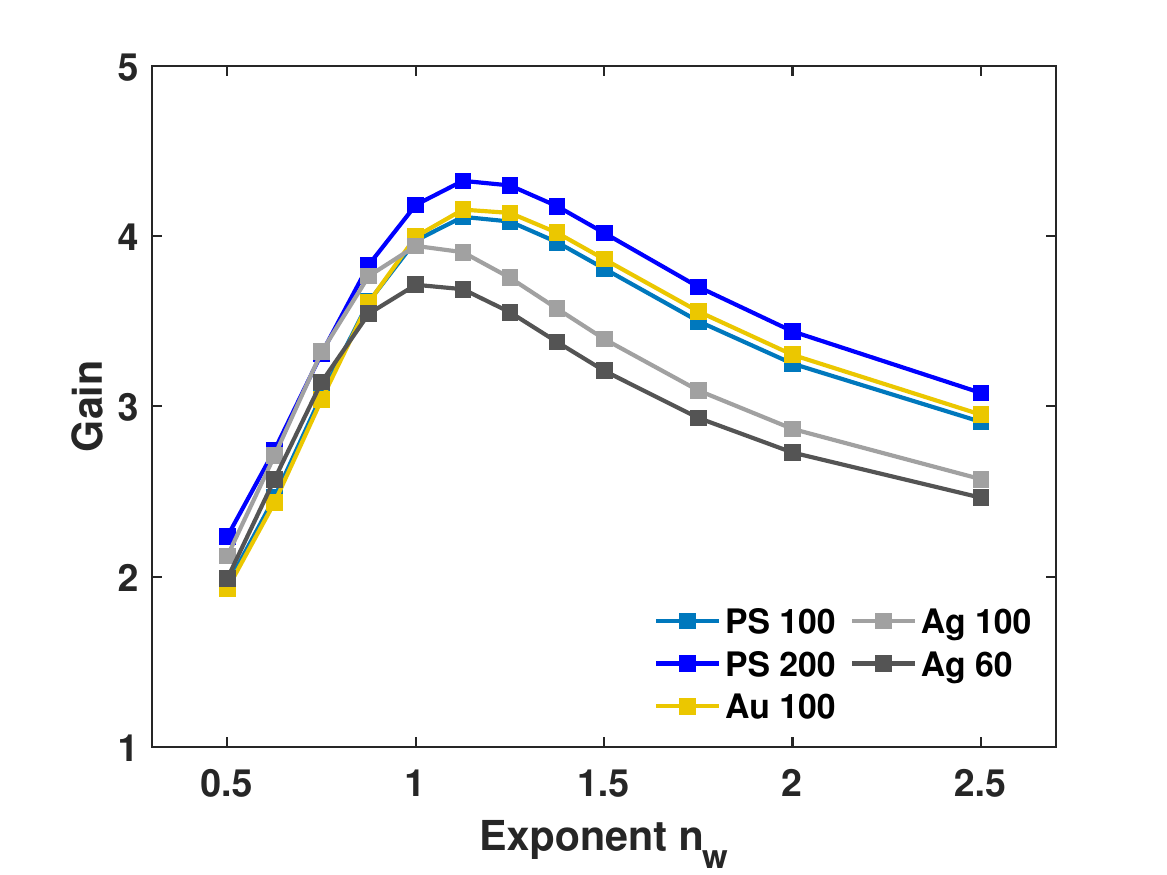}\label{fig:Simul_Fig_2_Diff_Particles_Gain}} \linebreak
	\sidesubfloat[]
	{\includegraphics[width=0.45\textwidth]{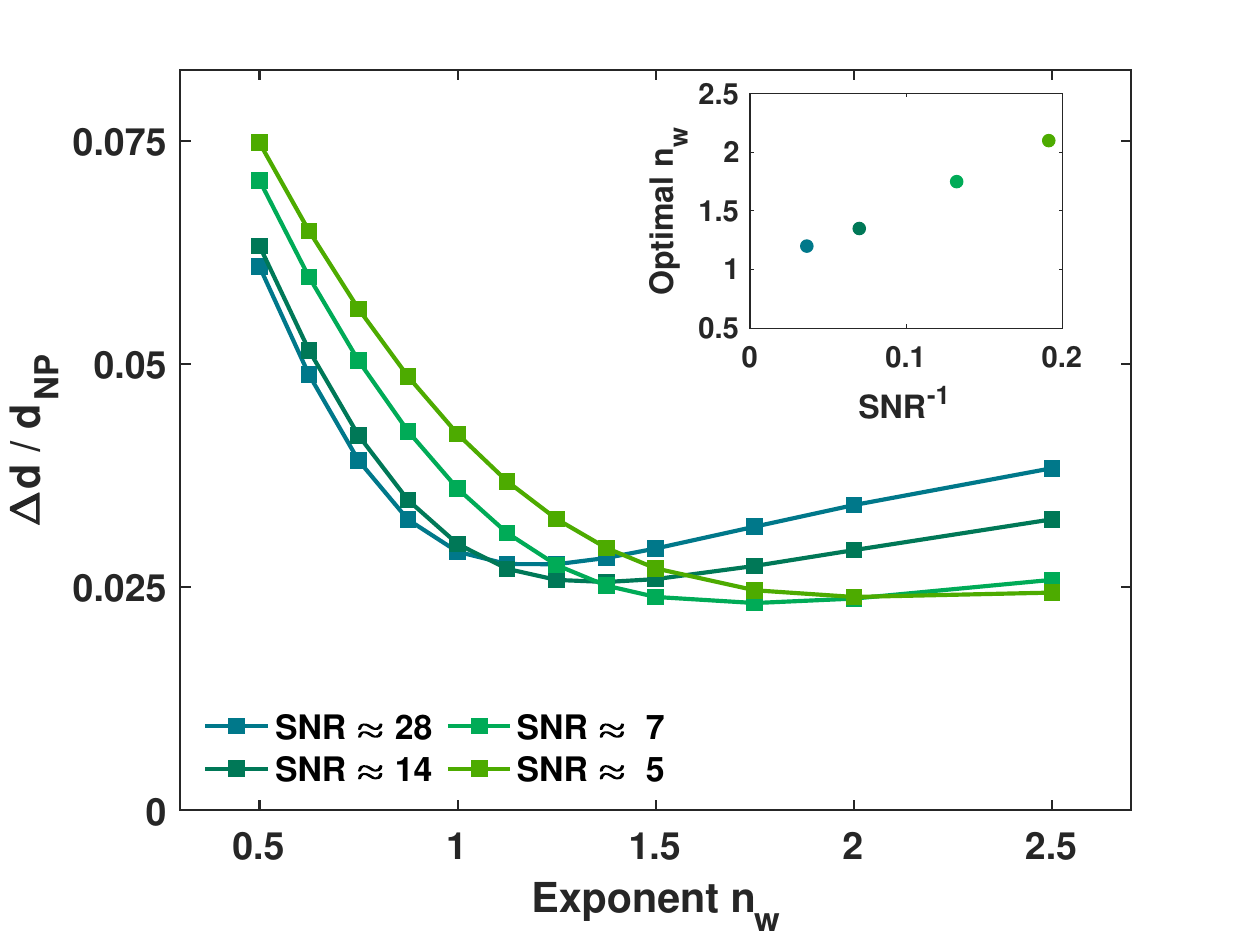}\label{fig:Simul_Fig_2_Diff_SNR}}
 	\sidesubfloat[]
	{\includegraphics[width=0.45\textwidth]{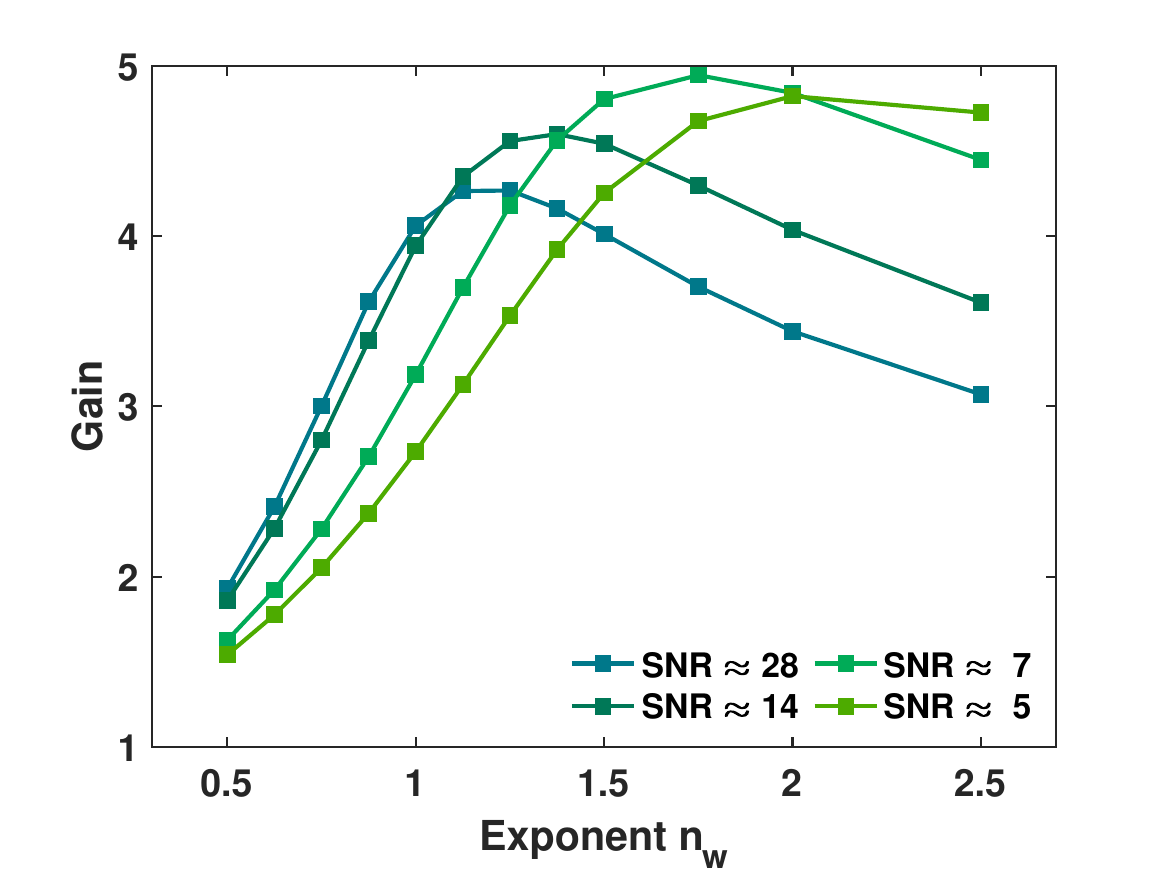}\label{fig:Simul_Fig_2_Diff_SNR_Gain}}
	\caption{\textbf{Analysis of the optimal exponent}. \textbf{(a, b)} Relative size difference and the gain of weighted NTA for different particles. Inset graph shows the optimal exponent for each particle. \textbf{(c, d)} Relative size difference and the gain of weighted NTA  of 100-nm PS at different SNR. Inset graph depicts a linear dependence of optimal exponent and SNR$^{-1}$.}
	\label{Simul_02}
\end{figure*}

Noteworthy, the optimal exponent is almost independent to the nature of particle (material, size) thanks to the Rytov field weighting approach, as illustrated in Figures \ref{fig:Simul_Fig_2_Diff_Particles} and \ref{fig:Simul_Fig_2_Diff_Particles_Gain}. While increasing photon shot noise (decreasing the SNR), the similarity between particles is less significant due to image noise. The averaging process is expected to be less important, leading to a shift of optimal exponent toward a higher value (Figure \ref{fig:Simul_Fig_2_Diff_SNR} and \ref{fig:Simul_Fig_2_Diff_SNR_Gain}). The optimal exponent $n_w$ is linearly dependant with the reciprocal of SNR (see inset of Figure \ref{fig:Simul_Fig_2_Diff_SNR}) and can thus be directly experimentally determined from the actual acquired phase and intensity images. Even at low SNR, the weighted fit remains useful with gain $>1$ when compared to classic NTA.\\

Simulation studies indicate that our approach presents a significant improvement of NP sizing for both mono- and poly-disperse solution of nanoparticles. In the subsequently step, experimental studies are performed in order to validate the method.

\subsection{Validation by experimental studies}

For experimental studies, intensity and phase images are acquired on a homemade microscope using a quadriwave lateral shearing interferometer \cite{Primot_1995,Primot_2000,Bon_2009}. Since we access to the full information of the scalar electromagnetic field, a moving particle can be numerically propagated at its focal plane \cite{MCN_2023_PhaseVirus}. A time-lapse averaging process is then applied after registering and cropping the image around the refocused particle. Details of the setup and the process are described in our previous work \cite{MCN_2023_PhaseVirus}. We determine the SNR at $\approx 10$ to $20$ for 100-nm PS NPs and the optimal exponent is estimated as $n_w = 1.25$. Since the actual size of each moving particle cannot be known exactly (each NP is smaller than the point spread function of the microscopy), the comparison between the size histograms of multiple tracked NPs of classic tracking and weighted tracking is used as a proof for the improvement of NTA. \\
\begin{figure*}[h!]    
	\centering
	\sidesubfloat[]
	{\includegraphics[width=0.45\textwidth]{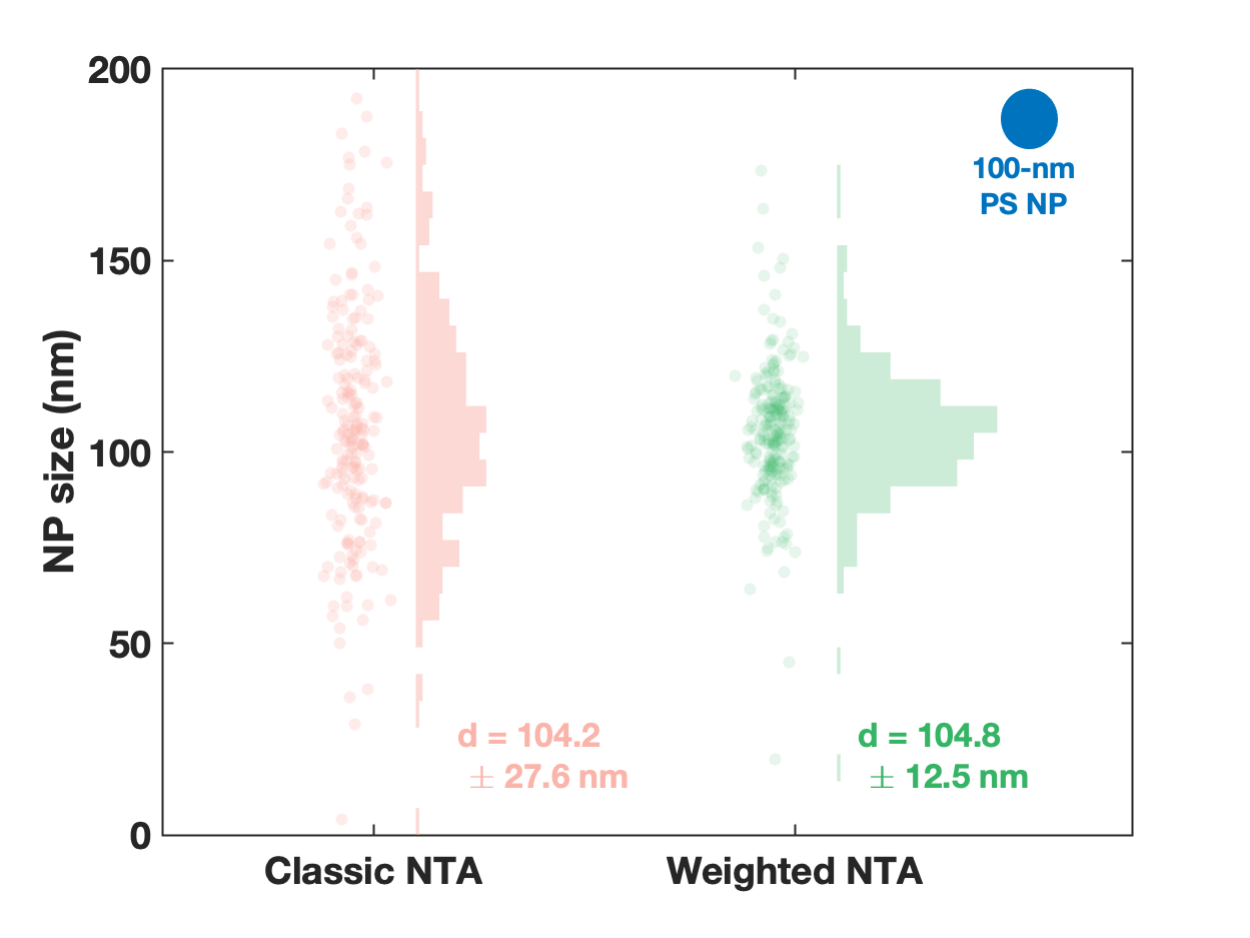}\label{fig:Exp_PS100}}
 	\sidesubfloat[]
	{\includegraphics[width=0.45\textwidth]{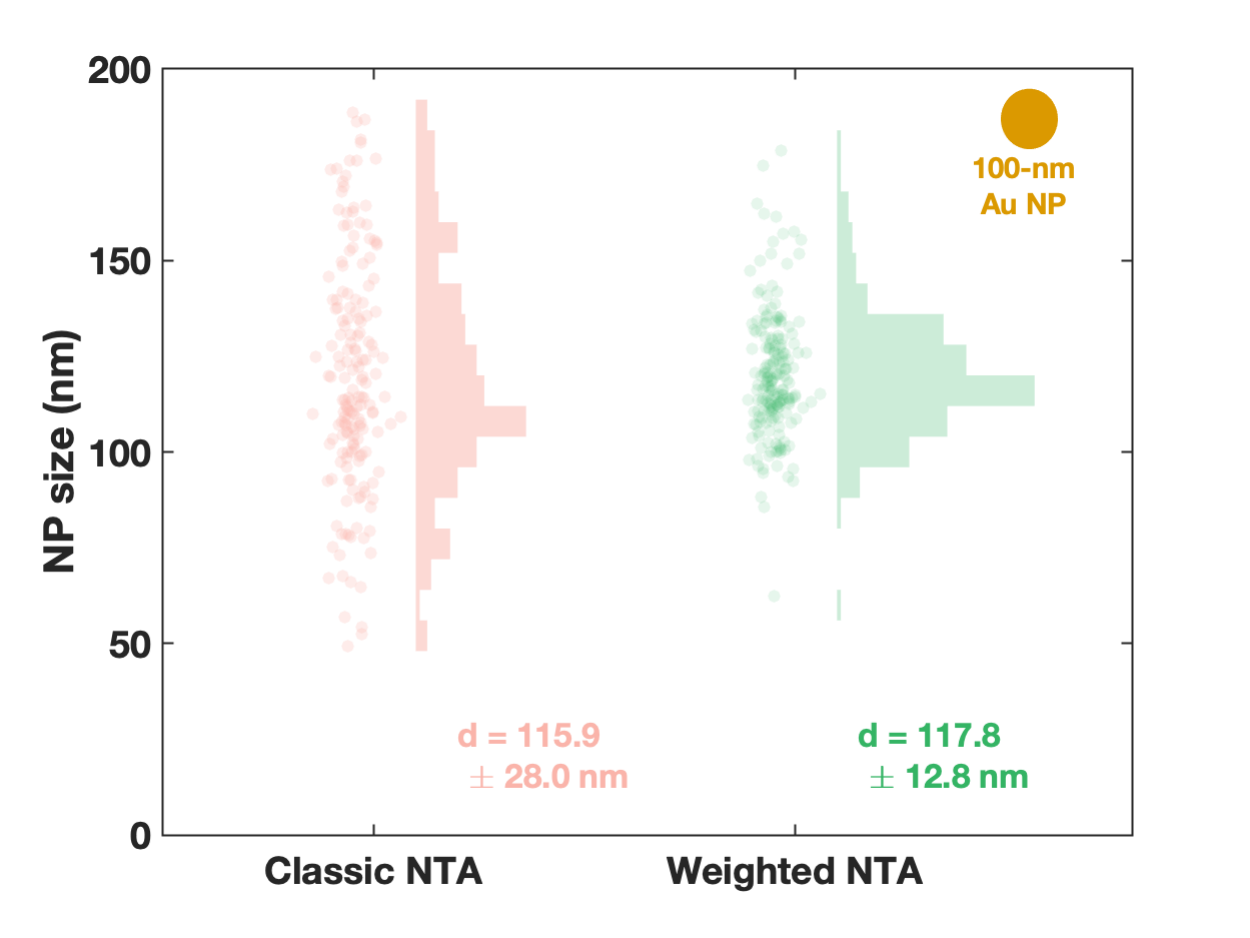}\label{fig:Exp_Au100}}\linebreak
 	\sidesubfloat[]
	{\includegraphics[width=0.45\textwidth]{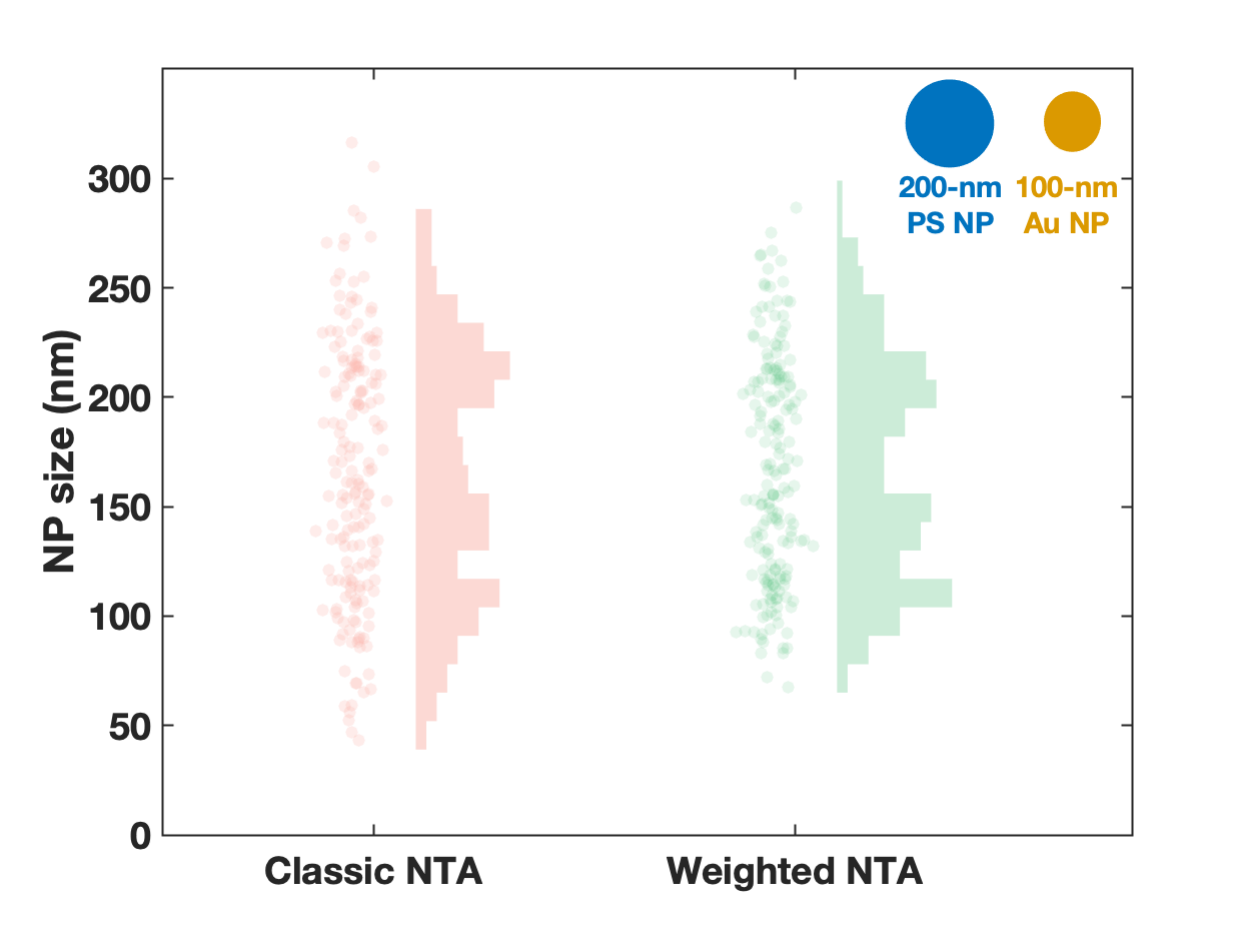}\label{fig:Exp_Mixture}} 
   	\sidesubfloat[]
	{\includegraphics[width=0.45\textwidth]{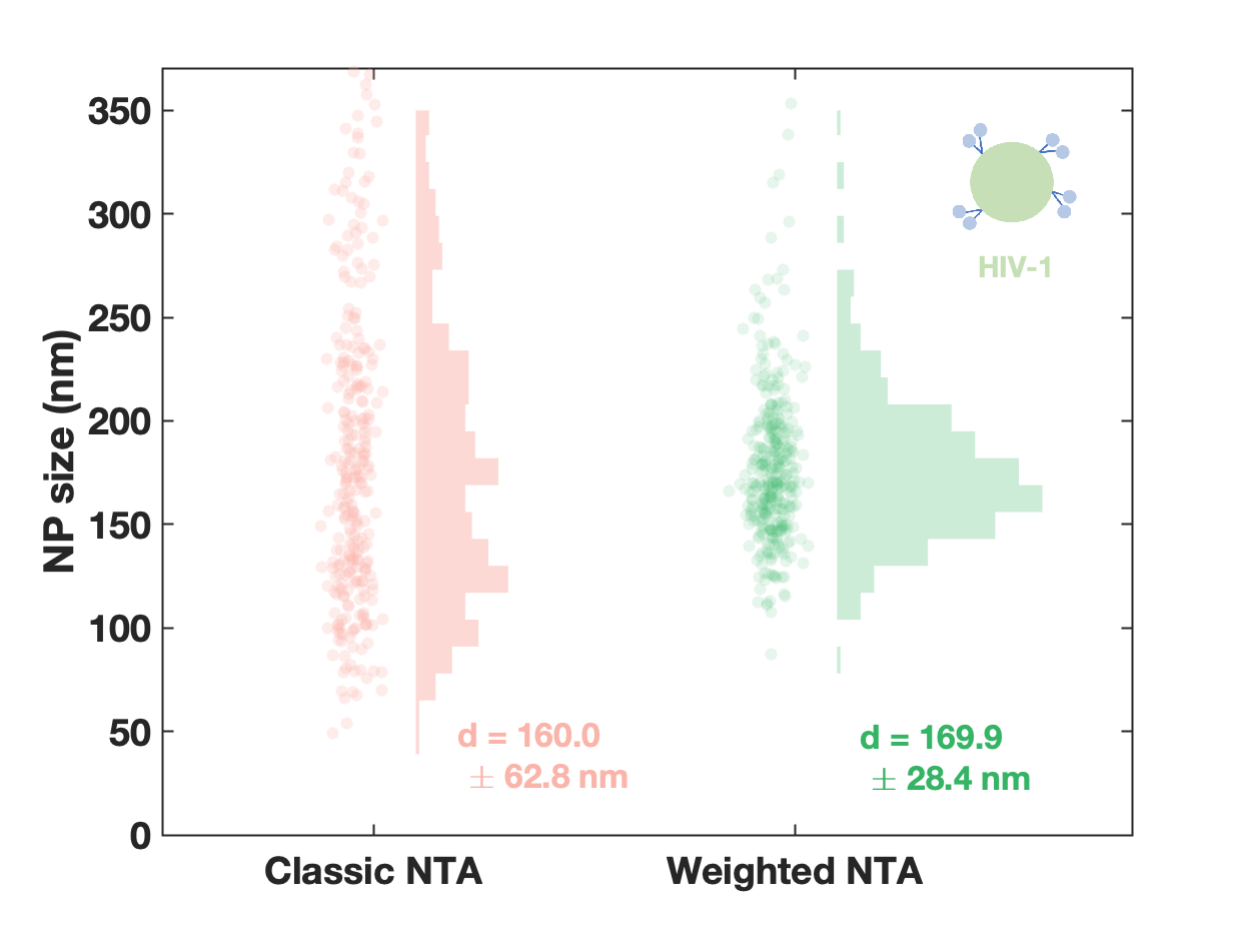}\label{fig:Exp_HIV}}
	\caption{\textbf{Experimental demonstration of MSD curve's weighted fit}. Comparison of size histograms, with and without weighted fit, for \textbf{(a)} mono-disperse 100-nm PS NP, \textbf{(b)} mono-disperse 100-nm Au NP, \textbf{(c)} poly-dispersed solution of 100-nm Au NP and 200-nm PS NP, and \textbf{(d)} infectious HIV-1 virus solution.}
	\label{Exp_demonstration}
\end{figure*}

Figure \ref{fig:Exp_PS100} shows the result of a calibrated PS solution (Duke Standards\textsuperscript{TM} 3K-100-001). Its properties are provided by the supplier (mean size of $102 \pm 3$ nm, S.D. of 7.6 nm, measured by transmission electron microscopy). The mean sizes are almost similar between classic NTA (104.2-nm) and our weighted approach (104.8-nm), and are in the confidence range provided by the supplier. Our MSD curve's weighted fit provides a better estimation of NP size dispersion in which the standard deviation is reduced by a factor 2, from 27.6-nm with classic NTA to 12.5-nm with the weighted approach (optimal exponent $n_w = 1.25$ determined from the images SNR), closer to the supplier information. The same observation is obtained for the solution of 100-nm Au NP (Sigma Aldrich), as shown in the Figure \ref{fig:Exp_Au100}. The size dispersion of Au NP is reduced by 2, and the mean size agrees to the hydrodynamic size provided by the supplier (118-nm). These results confirm the performance of our method for dielectric and metallic NPs.\\ 

Our method is then investigated for poly-disperse solution containing two classes of NPs: 200-nm PS and 100-nm Au NP (at 55\%:45\% molecular percentage, estimated from the dilution protocol). The two particles, one dielectric and the other metallic, are chosen so that their Rytov intensity are similar. This means that on raw phase and intensity images before numerical refocusing there signal are in absolute almost identical. Figure \ref{fig:Exp_Mixture} illustrates the size distribution of the mixture. With classic NTA it is challenging to separate the two classes and to properly distinguish that it was a mix and not a unique solution with very poly-disperse in size objects. The separation of the two classes is improved by the weighted NTA and we can clearly identify two populations, one centered at 200-nm and the other at 100-nm.\\

Our system is stable enough to carry out an experiment inside a biosafety cabinet of a level-3 confined laboratory (CEMIPAI, University of Montpellier). An example of the characterization of infectious HIV-1 (NL4-3) is illustrated in Figure \ref{fig:Exp_HIV}. Size distribution becomes sharper (2.2$\times$ gain factor) when weighted fit is applied allowing to determine the infectious HIV-1 size at $169.9 \pm 28.4$-nm, comparable to the measurement by cryo-electron microscopy ($145 \pm 25$-nm) \cite{Briggs_HIV_2003}. The difference is attributed to the outer-shell of glycoprotein gp160$^{Env}$ which participates in the hydrodynamic diameter measurement (tracking analysis) but not in the electron microscopy measurement. 

\subsection{Implementation in NP characterization}

Analysis of intensity and phase images is not restricted to single particle size determination only. The quantification of the scalar electromagnetic field can be used to characterize NPs by their refraction and absorption properties. For example, complex refractive index of individual NPs can be studied using quantitative phase imaging \cite{MCN_2023_PhaseVirus}, or  digital holography \cite{Midtvedt_2020}. Here, we presents an application of improved NTA in complex refractive index quantification by quantitative phase microscopy. In this approach, the complex refractive index $\tilde{n}$ is derived as an inverse proportion of the NP volume $V=\pi/6\cdot d^3$ \cite{MCN_2023_PhaseVirus}:
\begin{equation}
\tilde{n} = n_m - \frac{1}{ n_m \, V} \iint_S \Bigr[ \frac{1}{2} \, \text{Re}(E_{Ry}) + \text{Im}(E_{Ry}) \Bigr] \, dS ,
\label{eq:n_from_SIR}
\end{equation}

where, $\text{Re}(E_{Ry})$ and $\text{Im}(E_{Ry})$ real and imaginary part of Rytov field, $dS$ infinitesimally small surface element of the image $S$, $n_m$ refractive index of the surrounding medium. Therefore, the reduction of NP size error allows to increase the measurement accuracy since its influence is more important ($d^3$) than the other variables.\\ 
\begin{figure}[h!]    
	\centering
    \includegraphics[width=0.55\textwidth]{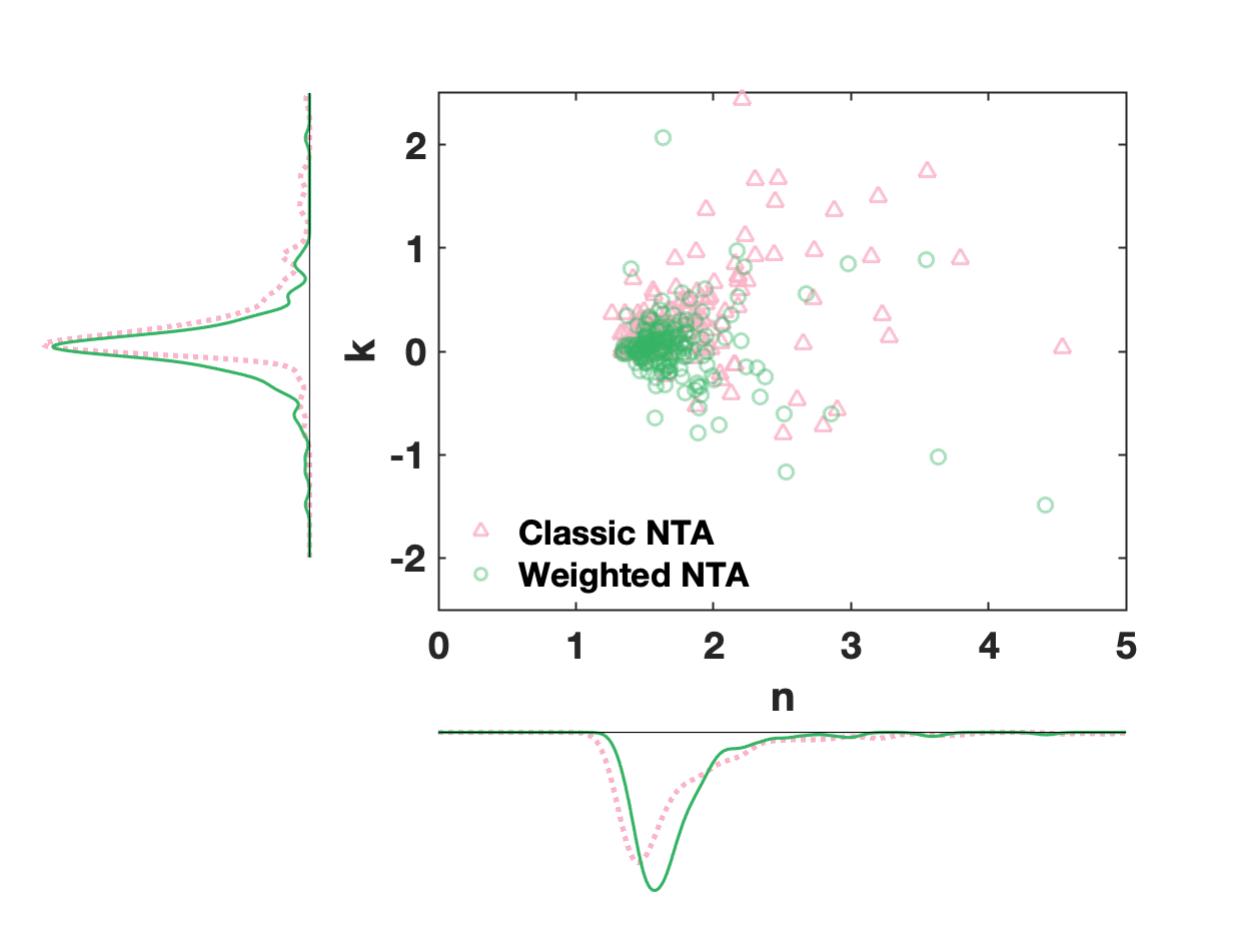}
	\caption{\textbf{Implementation to the quantification of refractive index}. Comparison of the measurement of complex refractive index of PS NP while using classic NTA and our improved NTA.}
	\label{Exp_RI_PS}
\end{figure}

Figure \ref{Exp_RI_PS} illustrates the results of complex refractive index ($n$ and $k$ are real and imaginary part) of 200-nm PS NPs, in both cases where NP size is derived from either classic NTA or our improved NTA. Our approach clearly reduces the measurement dispersion, from 0.35 to 0.16 for the real part. While using the improved NTA, the median of real part of refractive index of 200-nm PS NP is 1.64, closes to the literature ($n = 1.61$ in \cite{RI_PS}). The imaginary part is closes to 0, confirming the fact that PS is a transparent particle. The result of refractive index quantification justifies the effect of our improved weighted NTA in NP characterization.

\section{Conclusion}

In this work, we have presented a method to improve size evaluation from Brownian motion of single sub-resolved nanoparticles. Our approach is based on the weighted linear fit of MSD curves of all the tracked particles. Each weighted coefficient is automatically adjusted from the particle images of phase and intensity (image SNR). In our experimental condition, we have demonstrated that the dispersion of NP size can be reduced at least by a factor 2 compared to classic NTA. These results are illustrated for both mono- and poly-disperse solutions. The method can also be implemented in other NP characterization method, for example, in complex refractive index quantification, as demonstrated in this paper. Because the derivation of the refractive index involves NP size, the reduction of NP size error can enhance the measurement accuracy.  \\

Our approach is actually obtained thanks to the acquisition of intensity and phase images. We do believe that the method can be generalized for other imaging method, such as fluorescence microscopy or dark field microscopy. However, if only one observation is acquired (e.g. intensity of scattered light or intensity of fluorescent emitter), the confusion may occur in the case of poly-disperse solution, where two particles of huge difference in size but somehow interact with light in the same way. Therefore, the definition of similarity must be modified to cover the case.\\

\vspace{0.2cm}

\section*{Data availability statement}
The data cannot be made publicly available upon publication because they are not available in a format that is sufficiently accessible or reusable by other researchers. The data that support the findings of this study are available upon reasonable request from the authors.

\section*{Acknowledgments}
We thank Professor D. Muriaux and Dr S. Lyonnais at CEMIPAI, Université de Montpellier for the infectious virus sample, their support of setting up the experiment in their level-3 confined laboratory and the helpful discussions.

\section*{Funding}
This work was supported by CNRS. This project has received funding from the European Research Council (ERC) under the European Union’s Horizon 2020 research and innovation programme (grant agreement No. [848645]).

\section*{Competing interests}
All authors declare they have no competing interests. 

\newpage
\bibliographystyle{unsrt}
{\footnotesize
\bibliography{Reference_NP_Tracking} 
}

\end{document}